\documentclass[sigconf]{acmart}
\renewcommand\footnotetextcopyrightpermission[1]{}
\AtBeginDocument{%
  \providecommand\BibTeX{{%
    \normalfont B\kern-0.5em{\scshape i\kern-0.25em b}\kern-0.8em\TeX}}}

\acmConference[]{}{}{}

\usepackage{stmaryrd}  
\usepackage{ifthen}  
\usepackage{amsmath}  
\usepackage{booktabs}  
\usepackage{array}  
\usepackage[final]{listings}   
\usepackage{graphbox}

\newcommand*{\newmacro}[1]
  {\expandafter\newcommand\expandafter*\csname #1\endcsname}
\newcommand*{\renewmacro}[1]
  {\expandafter\renewcommand\expandafter*\csname #1\endcsname}
\newcommand*{\providemacro}[1]
  {\expandafter\providecommand\expandafter*\csname #1\endcsname}

\newcommand*{\usemacro}[1]
  {\csname #1\endcsname}

\let\newenvironment\newenvironment
\let\renewenvironment\renewenvironment
\let\provideenvironment\provideenvironment

\newmacro{isempty}[1]
  {\equal{#1}{}}

\newmacro{gobbleoneargument}[1]
  {}

\newmacro{newmathcommand}[3][]
  {\ifthenelse{\isempty{#1}}
    {\newmacro{#2}{#3}}
    {\newmacro{#2}{\usemacro{math#1}{#3}}}}
\newmacro{renewmathcommand}[3][]
  {\ifthenelse{\isempty{#1}}
    {\renewmacro{#2}{#3}}
    {\renewmacro{#2}{\usemacro{math#1}{#3}}}}

\newmacro{separate}
  {\medskip\noindent}

\providemacro{marginnote}
  {\marginpar}
\providemacro{smallcaps}
  {\textsc}
\providemacro{marginwidth}
  {\marginparwidth}

\newmacro{alert}[1]
  {\textbf{#1}}
\newmacro{divert}[1]
  {\textcolor{gray}{#1}}
\newmacro{enquote}[1]
  {``#1''}
\newmacro{fixme}[1]
  {\colorbox{yellow}{#1}\marginnote{\colorbox{yellow}{$\star$}}}
\newmacro{todo}[1]
  {\textcolor{red}{$\star$}\marginnote{\textcolor{red}{#1}}}
\newmacro{type}[1]
  {\texttt{#1}}

\newmacro{add}[1]
  {\textcolor{green}{#1}}
\newmacro{remove}[1]
  {\textcolor{red}{#1}}
\newmacro{change}[1]
  {\textcolor{orange}{#1}}
\newmacro{adjust}[2]
  {\remove{#1} \add{#2}}

\newcolumntype{L}{>{$}l<{$}}
\newcolumntype{C}{>{$}c<{$}}
\newcolumntype{R}{>{$}r<{$}}
\newcolumntype{T}{>{\ttfamily}l}
\newcolumntype{S}{>{\sffamily}l}

\newenvironment{block}
  {\begin{center}}
  {\end{center}}

\newmacro{newlogo}[2][]
  {\ifthenelse{\isempty{#1}}
     {\newlogoaux{#2}{\smallcaps{\lowercase{#2}}}}
     {\newlogoaux{#1}{#2}}}
\newmacro{newlogoaux}[2]
  {\newmacro{#1}{#2}}

\newmacro{newoperator}[1]
  {\newmathcommand[op]{#1}}
\newmacro{newkeyword}[2][]
  {\ifthenelse{\isempty{#1}}
    {\newoperator{#2}{\text{\normalfont\sffamily\bfseries #2}}}
    {\newoperator{#1}{\text{\normalfont\sffamily\bfseries #2}}}}
\newmacro{newvalue}[2][]
  {\ifthenelse{\isempty{#1}}
    {\newoperator{#2}{\text{\normalfont\sffamily #2}}}
    {\newoperator{#1}{\text{\normalfont\sffamily #2}}}}
\newmacro{newtype}[2][]
  {\ifthenelse{\isempty{#1}}
    {\newoperator{#2}{\text{\normalfont\sffamily\scshape #2}}}
    {\newoperator{#1}{\text{\normalfont\sffamily\scshape #2}}}}

\newmacro{obox}[2]
  {\makebox[0pt][l]{\ensuremath{#2}}\phantom{\ensuremath{#1}}}

\newmacro{highlight}[1]
  {\colorbox{lightgray}{\ensuremath{#1}}}

\newmacro{Quad}
  {\hspace{1.5em}}
\newmacro{Break}
  {\\[\smallskipamount]}

\newmacro{upon}
  {\genfrac{}{}{0pt}{0}}

\let\<\langle
\let\>\rangle

\newmathcommand[open]{llbrace}{\{\!|}
\newmathcommand[close]{rrbrace}{|\!\}}

\newmacro{set}[1]
  {\ensuremath{\{#1\}}}
\newmacro{tuple}[1]
  {\ensuremath{\<#1\>}}

\let\lt<
\let\gt>

\newmathcommand[bin]{pp}
  {+\!\!+}
\newmathcommand{Mid}
  {\;\mid\;}

\newmacro{powerset}[1]
  {\mathcal{P}(#1)}

\newmathcommand{n}{\underline{n}}

\newmathcommand[bb]  {NN}{N}
\newmathcommand[bb]  {ZZ}{Z}
\newmathcommand[bb]  {EE}{E}
\newmathcommand[bb]  {OO}{O}
\newmathcommand[bb]  {QQ}{A}
\newmathcommand[bb]  {RR}{R}
\newmathcommand[bb]  {CC}{C}
\newmathcommand[bb]  {HH}{H}

\newmathcommand[bb]  {LL}{L}
\newmathcommand[bb]  {UU}{U}
\newmathcommand[bb]  {BB}{B}
\renewmathcommand[bb]{SS}{S}

\let\to\rightarrow

\newmacro{hint}[1]
  {\quad\text{\{ #1 \}}}

\newmathcommand[op]{when}
  {\mathbf{when}}
\newmathcommand[op]{where}
  {\mathbf{where}}
\renewmathcommand[op]{and}
  {\mathbf{and}}
\newmathcommand[op]{otherwise}
  {\mathbf{otherwise}}
\newmathcommand[op]{impossible}
  {\mathrm{impossible}}

  {\begin{marginfigure}[#1]\equation}
  {\endequation\end{marginfigure}}

\newenvironment*{marginequation*}[1] 
  {\begin{marginfigure}[#1]\equation\nonumber}
  {\endequation\end{marginfigure}}

\newmacro{signature}[1]
  {\multicolumn{3}{@{}L@{}}{#1}}
\newmacro{inset}[1]
  {\multicolumn{3}{L}{\quad #1}}

  {\begin{block}\begin{tabular}{@{}rRCLl}}
  {\end{tabular}\end{block}}
\newenvironment*{grammar*}
  {\begin{block}\begin{tabular}{@{}RLl}}
  {\end{tabular}\end{block}}

\newmacro{placerule}[4]
  {\ensuremath{
    \upon
      {\text{\smallcaps{#1}}\hfill}
      {\dfrac{#2}{#3}\ #4}
  }}

\newmacro{newrule}[4]
  {\newmacro{#1}{\placerule{#1}{#2}{#3}{#4}}}
\newmacro{userule}
  {\usemacro}
\newmacro{refrule}[1]
  {\ifthenelse{\isundefined{#1}}
    {\GenericError{}{Rule `#1` is not defined}{}{}}
    {\textsc{#1}}}

\lstdefinestyle{common}
  {escapechar=|
  ,numbersep=-9pt 
  ,aboveskip=0pt
  ,belowskip=0pt
  }

\lstdefinestyle{natural}
  {style=common
  ,columns=fullflexible
  ,gobble=2
  ,breaklines=true
  ,breakatwhitespace=true
  ,literate=
    {<<}{{$\<$}}1
    {>>}{{$\>$}}1
    {->}{{$\to$\ }}2
  ,basicstyle={\sffamily}
  ,keywordstyle=[1]{\bfseries}
  ,keywordstyle=[2]{\scshape}
  ,keywordstyle=[3]{}
  ,emphstyle={\itshape}
  ,showstringspaces=false
  ,texcl=true
  ,mathescape=true
  ,xleftmargin=1\parindent
  }

\lstdefinestyle{flexible}
  {columns=flexible
  ,gobble=2
  ,fontadjust=true
  ,basicstyle={\ttfamily\small}
  ,commentstyle={\itshape}
  ,keywordstyle={\bfseries}
  ,emphstyle={\itshape}
  ,showstringspaces=false
  ,texcl=true
  ,mathescape=true
  ,xleftmargin=1\parindent
  }

\lstdefinestyle{literate}
  {style=natural
  ,literate=
    {\\}{{$\lambda$}}1
    {\\\$}{{\$}}1 
    {\\/}{{$\vee$}}1
    {/\\}{{$\wedge$}}1
    {A.}{{$\forall$}}1
    {E.}{{$\exist$}}1
    {->}{{$\rightarrow$ }}1
    {<-}{{$\leftarrow$}}1
    {==}{{$\equiv$\ }}1
    {/=}{{$\nequiv$\ }}1
    {<=}{{$\leq$}}1
    {>=}{{$\geq$}}1
    {>>=}{{>>=}}3 
    {\{|}{{$\{\!|\!$}}1
    {|\}}{{$\!|\!\}$}}1
    {\{|*|\}}{{$\{\!|\!\!\star\!\!|\!\}$}}3
  }

\lstdefinelanguage{tasks}
  {sensitive=true
  ,morekeywords=[1]{let,in,if,then,else,case,of,ref,assert,type}
  ,morekeywords=[2]{Bool,Int,String,Unit,List,Task, Passenger,Seat,Booking, Snack}
  ,moreemph={a,b,c,d,e,f,g,h,i,j,k,l,m,n,o,p,q,r,s,t,u,v,w,x,y,z as,bs,cs,ds,es,fs,gs,hs,is,js,ks,ls,ms,ns,os,ps,qs,rs,ss,ts,us,vs,ws,xs,ys,zs}
  ,morestring=[b]"
  ,morecomment=[l]--
  ,morecomment=[n]{\{-}{-\}}
  }[keywords,strings,comments]
\lstdefinestyle{tasks}
  {style=natural
  ,literate=
    {\\}{{$\lambda$}}1
    {<<}{{$\<$}}1
    {>>}{{$\>$ }}1
    {->}{{$\to$ }}1
    {==}{{$\equiv$ }}1
    {/=}{{$\nequiv$ }}1
    {<=}{{$\leq$ }}1
    {>=}{{$\geq$ }}1
    {*}{{$\times$ }}1
    {`elem`}{{$\in$ }}1
    {\\/}{{$\vee$ }}1
    {/\\}{{$\wedge$ }}1
    {>>=}{{$\Then$ }}1
    {>>?}{{$\Next$ }}1
    {<&>}{{$\And$ }}1
    {<|>}{{$\Or$ }}1
    {<?>}{{$\Xor$ }}1
    {++}{{$\pp$ }}1
    {edit}{{$\Edit$}}1
    {enter}{{$\Enter$}}1
    {update}{{$\Update$}}1
    {fail}{{$\Fail$ }}1
  }

\lstnewenvironment{TASK}[1][]
  {\lstset{language=tasks,style=tasks,#1}}
  {}
\newmacro{TS}[1][]
  {\lstinline[language=tasks,style=tasks,#1]}
\newmacro{includeTASK}[2][]
  {\lstinputlisting[language=tasks,style=tasks,#1]{#2}}

\let\phi\varphi

\newkeyword[IF]  {if}
\newkeyword[THEN]{then}
\newkeyword[ELSE]{else}

\newkeyword[Let]{let}
\newkeyword[In]{in}

\newmacro{If}[3]
  {\IF #1 \THEN #2 \ELSE #3}

\newmathcommand{unit}{\<\>}

\newvalue{True}
\newvalue{False}
\newvalue[Not]{not}

\newmacro{str}[1]
  {\text{``#1''}}

\newvalue[Map]{map}
\newvalue[Fst]{fst}
\newvalue[Snd]{snd}
\newvalue[Head]{head}
\newvalue[Tail]{tail}
\newvalue[Uniq]{uniq}
\newvalue[Len]{len}

\newtype{Unit}
\newtype{Bool}
\newtype{Nat}
\newtype{Int}
\newtype{String}
\newtype[Reference]{Ref}
\newtype{Task}
\newtype{Maybe}
\newtype{List}

\newtype{Euro}

\newmacro{TOPHAT}
  {$\widehat{\text{\smallcaps{top}}}$}
\newmacro{STOPHAT}
  {Symbolic $\widehat{\smallcaps{top}}$}

\let\And\relax
\newoperator{Then}  {\blacktriangleright}
\newoperator{Next}  {\vartriangleright}
\newoperator{And}   {\bowtie}
\newoperator{Or}    {\blacklozenge}
\newoperator{Xor}   {\lozenge}
\newoperator{Edit}  {\square}
\newoperator{View}  {\overline{\square}}
\newoperator{Enter} {\boxtimes}
\newoperator{Update}{\blacksquare}
\newoperator{Watch} {\overline{\blacksquare}}
\newoperator{Fail}  {\lightning}
\newoperator{At}    {@}

\newoperator{AndOr} {\DEPRECATED}

\newvalue[Left]   {L}
\newvalue[Right]  {R}

\newvalue[Empty]   {E}
\newvalue[Continue]{C}
\newvalue[Pick]    {P}

\newvalue[First]  {F}
\newvalue[Second] {S}
\newvalue[Here]   {H}

\newmathcommand[rel]{eval}
  {\;\downarrow\;}
\newmathcommand[rel]{stride}
  {\;\mapsto\;}
\newmathcommand[rel]{normalise}
  {\;\Downarrow\;}
\newmacro{handle}[1]
  {\mathrel{\;\xrightarrow{#1}\;}}
\newmacro{interact}[1]
  {\mathrel{\;\overset{#1}{\Longrightarrow}\;}}

  \newmathcommand[rel]{Leadsto}
   {\rotatebox[origin=c]{90}{\rotatebox[origin=c]{-90}{$\leadsto$}\!\rotatebox[origin=c]{-90}{$\leadsto$}}}

  \newmathcommand[rel]{simeval}
    {\;\rotatebox[origin=c]{-90}{$\leadsto$}\;}
  \newmathcommand[rel]{simstride}
    {\;\mapstochar\kern+0.08em\leadsto\;}
  \newmathcommand[rel]{simnormalise}
    {\;\rotatebox[origin=c]{-90}{$\Leadsto$}\;}
  \newmathcommand[rel]{simhandle}
    {\;\leadsto\;}
  \newmathcommand[rel]{siminteract}
    {\;\Leadsto\;}

\newmathcommand{simi}
   {\tilde{\imath}}
\newmathcommand{simt}
   {\tilde{t}}
\newmathcommand{sims}
   {\tilde{\sigma}}
\newmathcommand{sime}
   {\tilde{e}}
\newmathcommand{simv}
   {\tilde{v}}

\newmathcommand[it]{Simulate}
  {simulate}
\newmathcommand[it]{Again}
  {again}

\newmathcommand[cal]{Value}
  {V}
\newmathcommand[cal]{Inputs}
  {I}
\newmathcommand[cal]{Interface}
  {U}
\newmathcommand[cal]{Failing}
  {F}
\newmathcommand[cal]{Watching}
  {W}
\newmathcommand{Dirty}
  {\Delta}
\newmathcommand[cal]{UserInterface}
  {U}
\newmathcommand[cal]{Sat}
  {S}

\usepackage{xcolor}
\definecolor{codegreen}{rgb}{0,0.6,0}
\definecolor{codegray}{rgb}{0.5,0.5,0.5}
\definecolor{codepurple}{rgb}{0.58,0,0.82}
\definecolor{backcolour}{rgb}{0.95,0.95,0.92}

\lstdefinestyle{mystyle}{
    backgroundcolor=\color{backcolour},
    commentstyle=\color{codegreen},
    keywordstyle=\color{magenta},
    numberstyle=\tiny\color{codegray},
    stringstyle=\color{codepurple},
    basicstyle=\ttfamily\scriptsize,
    breakatwhitespace=false,
    breaklines=true,
    captionpos=b,
    keepspaces=true,
    numbers=left,
    numbersep=5pt,
    showspaces=false,
    showstringspaces=false,
    showtabs=false,
    tabsize=2,
    frame=lrtb,
    aboveskip=1.5em,
    belowskip=1.5em
}

\usepackage{subfig}
\usepackage[marginclue,inline]{fixme}
\fxsetup{status=draft}
\fxsetup{theme=color}
\fxsetup{envlayout=color}
\usepackage{url}

\begin{document}

\title{Creating Interactive Visualizations of TopHat Programs}

\author{Mark Gerarts}
\affiliation{%
  \institution{Open University}
  \city{Heerlen}
  \country{The Netherlands}
}
\email{mark.gerarts@gmail.com}

\author{Marc de Hoog}
\affiliation{%
  \institution{Open University}
  \city{Heerlen}
  \country{The Netherlands}
}
\email{mla.dehoog@studie.ou.nl}

\author{Nico Naus}
\affiliation{%
  \institution{Virginia Tech}
  \city{Blacksburg, VA}
  \country{United States}
}
\email{niconaus@vt.edu}

\author{Tim Steenvoorden}
\affiliation{%
  \institution{Open University}
  \city{Heerlen}
  \country{The Netherlands}
}
\email{tim.steenvoorden@ou.nl}

\renewcommand{\shortauthors}{Gerarts and de Hoog, et al.}

\begin{abstract}


Many companies and institutions have automated their business process in
workflow management software. The novel programming paradigm Task-Oriented
Programming (TOP) provides an abstraction for such software. The largest
framework based on TOP, iTasks, has been used to develop real-world software.

Workflow software often includes critical systems. 
In such cases it is important to reason over the software to ascertain its correctness.
The lack of a formal iTasks semantics makes it unsuitable for formal reasoning.
To this end TopHat has been developed as a TOP language with a formal semantics. 
However, TopHat lacks a graphical user interface (GUI), making it harder to develop practical TopHat systems.

In this paper we present TopHat UI.
By combining an existing server framework and user interface framework, we have
developed a fully functioning proof of concept implementation in Haskell, on top of TopHat's semantics.
We show that implementing a TOP framework is possible using a different host language than iTasks uses.
None of TopHat's formal properties have been compromised, since the UI framework is completely separate from TopHat.
We run several example programs and evaluate their generated GUI.
Having such a system improves the quality and verifiability of TOP software in general.


\end{abstract}

\keywords{task oriented programming, user interface, functional programming}

\maketitle
\pagestyle{plain}


\section{Introduction}

Workflow software is present in most businesses and institutions
nowadays. From health care and first responders, to commerce and industrial
processes. Businesses use workflow software to streamline their processes,
increase efficiency and reduce costs. In these sectors, reliability of software
is crucial.

Previous research into workflow software in the functional
programming community aimed to improve reliability, while at the same time
reducing the effort of development. This led to the development of
Task-Oriented Programming (TOP), a programming paradigm that aims to facilitate
working with multiple people towards a shared goal over the internet. TOP
separates the \emph{what} from the \emph{how}. This separation allows
programmers to focus on the work that has to be done (\emph{what}) instead of
paying attention to design issues, implementation details, operating system
limitations, and environment requirements
(\emph{how})~\cite{achten2015top,plasmeijer2012taskoriented}.

\textit{iTasks}~\cite{achten2015top}, implemented in the functional programming language
Clean~\cite{clean}, is the main TOP framework and has been around for a long time.
iTasks has been used to create real-world applications, such as an incident
coordination tool for the Dutch coast guard~\cite{lijnse2012incidone}. While
this proves its practical usability, iTasks lacks in formalization.
The iTasks semantics are given by its implementation, making it much harder to formally reason about iTasks programs.
Previous attempts to mitigate this issue by some of iTasks' creators involved developing a separate iTasks semantics, which allowed them to perform model-based testing, but no formal verification~\cite{DBLP:conf/ifl/KoopmanPA08}.
Formal program verification is a very powerful tool to ensure the correctness of critical
software, like the incident coordination tool.
TopHat is a Domain-Specific Language (DSL) that paves the way to formally reason about task-oriented
programs~\cite{steenvoorden2019tophat}, by defining a formal TOP semantics.
These semantics have been implemented in Haskell and Idris~\footnote{\url{https://github.com/timjs/tophat-proofs}}.
Idris is a programming language that features dependent types and a totality checker, which is used to prove properties of TopHat programs.

\subsection*{Motivation}
In this paper, we develop an interactive UI for TopHat.
Before the development of TopHat, it was the case that iTasks, TOP and Clean were tied together very strongly.
Previous research even suggests that certain specific Clean features are essential to the implementation of TOP\cite{plasmeijer2012taskoriented}:
uniqueness typing, data generic programming, dynamics~\cite{DBLP:conf/ifl/VervoortP02} and a sophisticated backend using interpreted ABC bytecode on clients \cite{Oortgiese2017distributed}, to name a few.
We want to show that none of those features are essential in implementing a TOP framework with a GUI.
On top of that, we want to demonstrate that this can be achieved without making any changes to the TopHat language and its implementation in Haskell.

We expect this work to bring TOP to a bigger audience.
The current Clean user base is quite small.
Haskell is being used in production code, has a huge number of packages available online and an active community.
Task-oriented programming could benefit from being ported to Haskell, making it available to a large community of both developers and researchers.
Developing an interactive UI for TopHat brings this one step closer.

Motivated by the above, this paper presents a prototype framework written on top of TopHat's
Haskell implementation that is able to create interactive graphical user interfaces of
TopHat programs.



\subsection*{Structure}

The remainder of this paper is structured as follows: we first provide some
background about TOP, including iTasks and TopHat in Section~\ref{sec:top}.
Section~\ref{sec:tophat-user-interface} introduces our TopHat UI prototype.
Section~\ref{sec:examples} demonstrates the capabilities of our framework, including formal reasoning, using several example TopHat programs.
We highlight related work in Section~\ref{sec:related-work}.
Section~\ref{sec:discussion} reflects on the goals and research questions outlined above.
Section~\ref{sec:conclusion} concludes.


\section{Task-Oriented Programming}
\label{sec:top}

This paper builds upon previous TOP research~\cite{plasmeijer2012taskoriented,achten2015top,steenvoorden2019tophat}.
In this section we describe the basic idea of TOP and two TOP
implementations: iTasks and TopHat.

\subsection{Task-Oriented Programming}

TOP is centered around the concept of \textit{tasks}, which specify the work a
user or system has to perform with a high level of abstraction.
The smallest possible task represents the smallest amount of
work a user or system can perform~\cite{plasmeijer2012taskoriented}.

Combining small tasks allows creating large and complex applications
using simple building blocks. Tasks can be combined using combinators:
they can be executed sequentially, in parallel, or conditionally.
These combinators closely resemble how collaboration happens in real life.

TOP aims to facilitate collaborating with multiple people towards a shared goal, over the internet.
Creating complex applications is further facilitated because
tasks are first-class citizens: they can be used as input of functions, they can
be returned from them, and tasks can contain other tasks as value.

Tasks are interactive and input-driven. When a task receives input it is
reevaluated and results in a new task. A task's value can be observed at all
times. Tasks can share information with each other, either directly through
shared data stores, or by passing task values to continuations.

TOP itself focuses on the domain logic, with tasks providing merely a description of
the work that has to be performed. It is left up to a TOP framework to do the
heavy lifting, such as generating the user interface, storing and handling data,
setting up a web server, and authenticating users.
\textit{iTasks}~\cite{achten2015top} is such a
framework, implemented in the functional programming language
Clean~\cite{clean}. An example of a basic task in iTasks is presented in
Listing~\ref{lst:task-itasks}. Developers only have to specify that they
want the user to enter some information. Passing this task description to iTasks generates
an application that prompts the user for their name.

\begin{lstlisting}[language=Haskell,caption={A simple task prompting the user for their name (Clean)},label={lst:task-itasks}]
enterName :: Task String
enterName = Hint "What is your name?" @>> enterInformation []\end{lstlisting}

The TOP paradigm provides an abstraction over workflow software. Instead of
having to write a server, database, user interfaces, etc, programmers just
define what needs to be done. The complete application is then derived from this
specification by the TOP framework. TOP is usually embedded in pure functional programming.

To summarize, TOP is made up of the following three core concepts:

\begin{description}
      \item[Tasks] that describe the work that has to be performed, providing
            an abstraction that separates the \textit{what} from the
            \textit{how}~\cite{achten2015top}.
      \item[Composition] of tasks through combinators, allowing the
            creation of arbitrarily large tasks.
      \item[Data] that is being passed between tasks sequentially and globally.
\end{description}


\subsection{iTasks}
\label{sec:context:itasks}

iTasks~\cite{plasmeijer2007itasks} is a TOP framework that uses
Clean~\cite{brus1987clean} as its host language. It supplements Clean with a set
of combinators, model types, and algorithms that allow the construction of
task-oriented programs.

\begin{figure}[t]
      \centering
      \includegraphics[width=0.47\linewidth]{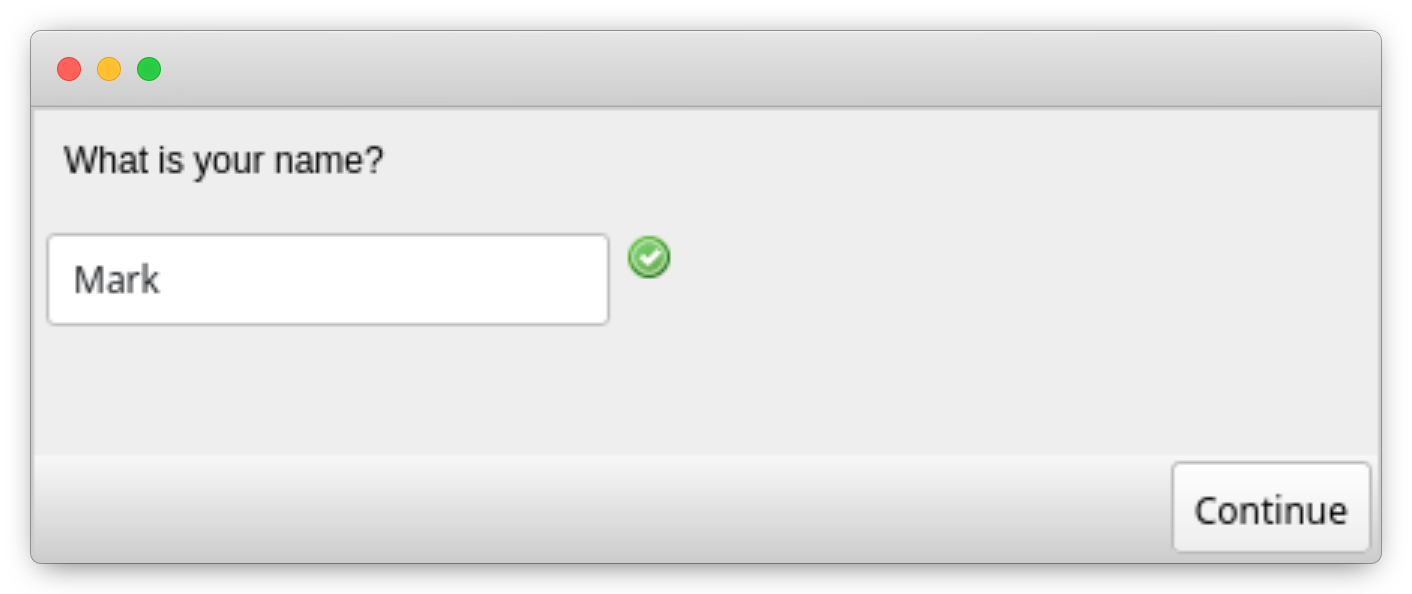}
      \includegraphics[width=0.47\linewidth]{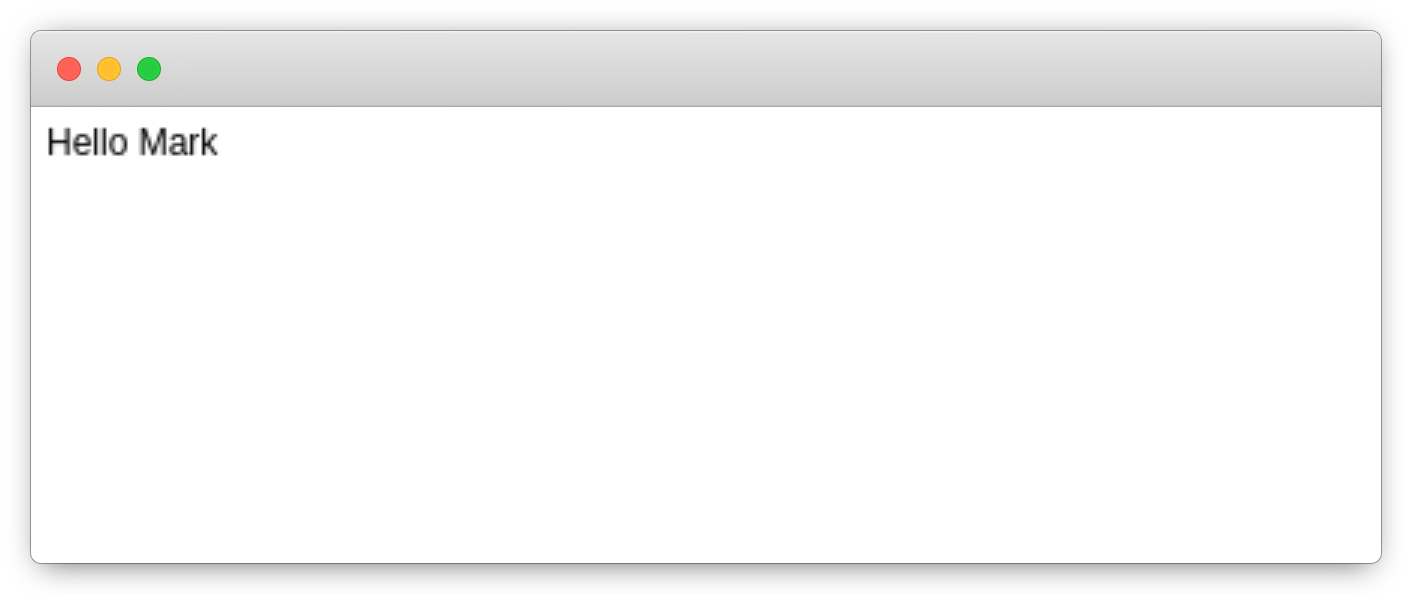}
      \captionsetup{type=figure}
      \captionof{figure}{Entering your name (left) and the result after pressing continue (right)}
      \label{fig:itasks-greet}
\end{figure}

An example of a basic task was given in Listing~\ref{lst:task-itasks}. iTasks
will automatically generate an entire application for this task. It uses
generics to deduce that a task of type \texttt{String} requires a text input
field. In Listing~\ref{lst:itasks-greet} we
combine the task with a view task using a sequential step combinator. A
user has to enter their name and is greeted by the program after stepping to
the next task. Figure~\ref{fig:itasks-greet} shows how these steps would look in
iTasks.

\begin{lstlisting}[language=Haskell,caption={Combining two tasks with a step combinator (Clean)},label={lst:itasks-greet}]
greet :: Task String
greet = enterName >>!
          \result -> viewInformation [] ("Hello " +++ result)\end{lstlisting}

iTasks is a work in progress, receiving constant updates and improvements. For
example, a recent addition is the usage of a distributed, dynamic
infrastructure~\cite{Oortgiese2017distributed}. iTasks has formed the basis of
further research as well. Tonic~\cite{stutterheim2014tonic} facilitates the
subject for non-technical people by providing graphical blueprints of iTasks
specifications. It also provides a way to monitor the process while end users
are interacting with the application~\cite{stutterheim2019static}. iTasks acted
as the starting point for research into declarative user interfaces, first for
SVG images~\cite{achten2014itasks} and later as a generalized
solution~\cite{achten2016layout}.

\subsection{TopHat}
\label{sec:tophat}

When software is used in critical applications, it is important that its
behavior can be verified and formally reasoned about. iTasks is primarily
focused on practical applicability, and therefore lacks this formalization.
Testing an iTasks application is time consuming and often incomplete because of
the many different execution paths.

TopHat~\cite{steenvoorden2019tophat} distills TOP's core features
to provide a way to reason about task-oriented programs. By employing
symbolic execution it is possible to formally verify TopHat
programs~\cite{naus2019symbolic}. Symbolic execution has also been used to
provide end-users of tasks with additional feedback~\cite{naus2020generating}.

Our work is based on TopHat's Haskell implementation.
Listing~\ref{lst:tophat-greet} gives the TopHat implementation of the example introduced in Section~\ref{sec:context:itasks}.
Similar to the iTasks code, this task uses a step combinator to ask a user their name and subsequently greet them.

\begin{lstlisting}[language=Haskell,caption={A TopHat task that greets the user (Haskell)},label={lst:tophat-greet}]
greet :: Task h String
greet = enter >>? \result -> view ("Hello " ++ result)\end{lstlisting}

TopHat contains the following set of tasks and combinators:

\begin{description}
      \item[Editors] model user interaction.
            They are typed containers
            that are either empty or hold a value.
            TopHat contains different kinds of editors:
            \begin{description}
                  \item[Update] contains a predefined value.
                  \item[View] is an editor with a view-only value.
                  \item[Enter] is an editor that is initially empty. Filling it transforms it into an Update editor.
                  \item[Watch] displays the value of a shared data store.
                  \item[Change] is an editor that allows to change the value of a shared data store.
            \end{description}
      \item[Done and Fail] are success and failure end tasks.
      \item[Pair] combines two tasks (parallel-and).
      \item[Choose] makes a choice between two tasks (parallel-or).
      \item[Step] sequentially moves from one task to another.
      \item[Share] creates a shared data store.
      \item[Assign] assigns a value to a reference in a shared data store. 
\end{description}

\subsection{Formal reasoning}
\label{sec:top:formal}

iTasks defines tasks as a ``state transforming function that reacts to an event, rewrites itself to a reduct and accumulates responses to users''~\cite{plasmeijer2012taskoriented}.
For combinators, iTasks takes the swiss-army-knife-approach.
It defines two combinators that perform a multitude of actions.
From these combinators, more simple ones can be constructed.
For example, the \texttt{>>*} combinator performs sequential composition, allows the user to choose from a list of tasks, allows automatic progressing tasks, guarded tasks, and stepping on exception.
Its definition in the latest version of iTasks is about 100 lines of Clean code, relying on many custom functions~\footnote{\url{https://gitlab.com/clean-and-itasks/itasks-sdk/-/blob/master/Libraries/iTasks/WF/Combinators/Core.icl}}.
While iTasks is certainly an impressive engineering accomplishment, it is unfit for formal reasoning.

TopHat on the other hand defines tasks as a simple datatype, with three base cases and a small number of simple combinators~\cite{steenvoorden2019tophat}.
The TopHat framework takes care of handling events, rewriting and task rendering.
The formal TopHat semantics fits on a single page, and is largely straightforward.

To demonstrate the formal reasoning capabilities of TopHat, a symbolic execution semantics has been developed~\cite{naus2019symbolic}.
For space reasons, we will refrain from repeating syntax and semantics here, but will revisit an example, to use thoughout this paper.

\lstset{emph={invoiceDate,date,confirmed,invoiceAmount,approved}}
\begin{TASK}[float=ht
            ,numbers=right
            ,caption=Subsidy request and approval workflow at the Dutch tax office.
            ,label=lst:tax
            ]
  let today = $\text{25 Sept 2020}$ in
  let provideDocuments = enter Amount <&> enter Date in
  let companyConfirm = edit True <?> edit False in
  let officerApprove = \ invoiceDate. \ date. \ confirmed.
    edit False <?> if (date - invoiceDate < 365 /\ confirmed) |\label{lst:tax:officer-approve-def}|
      then edit True
      else fail in
  provideDocuments <&> companyConfirm >>= |\label{lst:tax:documents-and-company-confirm}|
    \ <<<<invoiceAmount, invoiceDate>>, confirmed>>.
  officerApprove invoiceDate today confirmed >>= \ approved.|\label{lst:tax:officer-approve}|
  let subsidyAmount = if approved
    then min 600 (invoiceAmount / 10) else 0 in
  edit <<subsidyAmount, approved, confirmed, invoiceDate, today>>|\label{lst:tax:result}|
\end{TASK}

Listing~\ref{lst:tax} provides the code for a small example task, implementing the process of applying for a tax subsidy.
This example was inspired by a collaboration with the Dutch Tax office.
The user gets asked to provide documents to back up their tax subsidy request for solar panel installation (line 2).
The installation company has to confirm that they installed the panels (line 3), which can be done in parallel (line 8).
Finally, a tax officer can either approve or deny the request (line 4), depending on certain conditions (line 5).
After the task has been completed, the subsidy amount is being calculated (line 12), and the details are returned in a view (line 13).

For this task, symbolic execution allowed the authors to prove correctness properties over the code, such as functional correctness.
In Section~\ref{sec:tax} we will take a look at generating a UI using the framework presented in the coming section.


\section{TopHat UI Framework}
\label{sec:tophat-user-interface}

In this section we describe our prototype TOP UI framework, which is a
proof-of-concept and not a fully fledged TOP framework. Our application supports
TopHat tasks as mentioned in Section~\ref{sec:tophat}. We limit
ourselves to a select number of datatypes: only integers, booleans, and strings
are supported. Advanced framework features such as multi-user support are out of
scope as well.
We will reflect on this in Section~\ref{sec:conclusion}.
The framework is written in Haskell, and we use the following extensions.
\begin{description}
  \item [OverloadedLists] to allow for a more convenient HashMap notation.
  \item [OverloadedStrings] to allow for a more convenient way of using Text.
  \item [PackageImport and NoImplicitPrelude] to deal with the fact that TopHat defines its own Prelude.
\end{description}
All source code is published on GitHub\footnote{\url{https://github.com/mark-gerarts/ou-afstuderen-artefact}},
along with the examples described below. 

Key to our approach is that we leave the task specification of TopHat untouched.
This preserves the nice formal properties for which TopHat has been developed in the first place.
The prototype UI framework completely relies on the TopHat semantics for handling input and rewriting tasks.
The responsibility of the UI framework is to render the task in a web browser, and hand off input that comes in from the user to the TopHat semantics.

\begin{figure}[t]
    \centering
    \includegraphics[width=\linewidth]{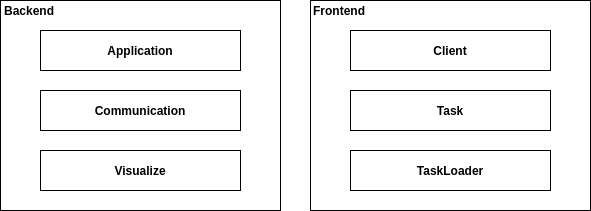}
    \captionsetup{type=figure}
    \captionof{figure}{Architecture. Each box represents a main module.}
    \label{fig:architecture}
\end{figure}

The prototype framework is architecturally separated in two parts: the backend
and the frontend. Figure~\ref{fig:architecture} shows the main modules of each
part. The backend is responsible for initializing tasks and handling
communication with TopHat. The frontend renders tasks and allows the user to
interact with them. After a comparative study of existing web server and UI
frameworks~\cite{markmarc2021}, we have selected Servant~\cite{servant} as our webserver and
Halogen~\cite{purescripthalogen} for the UI. Other options are discussed in the
Section~\ref{sec:related-work}.
Section~\ref{sec:artefact:communication} illustrates the communication
between frontend and backend. Section~\ref{sec:artefact:backend} explains
the working of the backend and the frontend is discussed in Section~\ref{sec:artefact:frontend}.

\begin{figure}[t]
    \centering
    \includegraphics[width=0.75\linewidth]{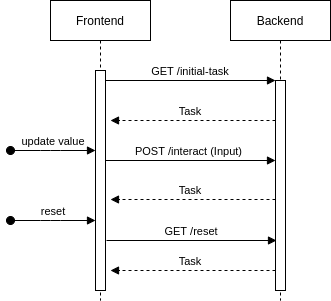}
    \captionsetup{type=figure}
    \captionof{figure}{Communication between frontend and backend. Sequence diagram that displays requests (solid arrows) and responses (dashed arrows). \texttt{update value} and \texttt{reset} are user actions. Task and Input are JSON objects.}
    \label{fig:frontendBackendSSD}
\end{figure}

\subsection{Communication between backend and frontend}
\label{sec:artefact:communication}

Figure \ref{fig:frontendBackendSSD} shows the communication between frontend and
backend. The frontend first requests the initial task, which the backend returns
using a JSON representation of this task. A user can now interact with the
system. In this example, the user updates a value. The frontend sends the input
as JSON to the backend, and the backend responds with the updated task. This
step can be repeated as necessary. In this case, the user resets the
application, which results in the backend resetting back to the initial task.

The frontend is written in PureScript and the backend in Haskell. We choose JSON
as data interchange format, because JSON allows custom data structures, it is
easy to use, and both backend and frontend support JSON out-of-the-box.

\subsection{Backend}
\label{sec:artefact:backend}

The backend is written in Haskell, using Servant~\cite{servant} as the web
server.
It has three main responsibilities, which is reflected in its module structure, shown in Figure~\ref{fig:architecture}:

\begin{enumerate}
    \item The Application module loads the application, defines the web
          server and configures the handlers.
    \item The Communication module handles JSON conversion, both encoding tasks
          to their JSON representation and decoding user input.
    \item The Visualize module is intended for the end user. It exposes
          functions to start the framework, which is demonstrated in
          Listing~\ref{sec:artefact:usage}.
\end{enumerate}

\begin{lstlisting}[caption={Starting the framework (Haskell)},label={sec:artefact:usage},language=Haskell]
import Task (Task, enter, view, (>>?))
import Visualize (visualizeTask)

main :: IO ()
main = visualizeTask greet

greet :: Task h String
greet = enter >>? \result -> view ("Hello " ++ result)
\end{lstlisting}

\paragraph{Application module}

We create an abstract web application (WAI-application) in the Application
module (see the \texttt{application} function in Listing
\ref{sec:artefact:lst:application}). We define the endpoints, the request and
the response formats. For example, see the \texttt{TaskAPI} in Listing
\ref{sec:artefact:lst:application}. The \texttt{server} function provides
handlers to serve the initial task, to handle interaction with the frontend and
to perform a reset. The remainder of the module consists of functions that
expose functionality of TopHat: initializing tasks, deconstructing tasks in a
representation that can be sent to the frontend, and interacting with tasks. We
have only added key signatures to Listing \ref{sec:artefact:lst:application}.

\begin{minipage}{\linewidth}
\begin{lstlisting}[caption={Application module (Haskell)},label={sec:artefact:lst:application},language=Haskell]
module Application (application, State (..)) where

data State h t = State
    { currentTask :: TVar (Task RealWorld t),
      initialised :: Bool,
      originalTask :: Task RealWorld t
    }

type TaskAPI =
    "initial-task" :> Get '[JSON] TaskDescription
    :<|> "interact"
        :> ReqBody '[JSON] JsonInput :> Post '[JSON] TaskDescription
    :<|> "reset" :> Get '[JSON] TaskDescription

type StaticAPI = Get '[HTML] RawHtml :<|> Raw
type API = TaskAPI :<|> StaticAPI

interactIO :: Input Concrete -> Task RealWorld a -> IO (Task RealWorld a)
initialiseIO :: Task RealWorld a -> IO (Task RealWorld a)
describeIO :: Task RealWorld a -> IO TaskDescription

server :: ToJSON t => State h t -> ServerT API (AppM h t)

application :: ToJSON t => State h t -> Application\end{lstlisting}
\end{minipage}

\paragraph{Communication module}

In Listing \ref{sec:artefact:lst:communication} we show the core of the
communication module. We introduce a new datatype, \texttt{TaskDescription},
that holds all data we need to render a task: the task itself
(\texttt{JsonTask}) and its possible inputs (\texttt{InputDescription}), along
with the \texttt{describe} function that extracts this data from a TopHat task.
User input, which is sent back and forth from the client to the server, is
defined in \texttt{JsonInput}.

\begin{lstlisting}[caption={Communication module (Haskell)},label={sec:artefact:lst:communication},language=Haskell]
module Communication (JsonTask (..), TaskDescription (..), describe) where

type JsonTask = Value

type InputDescriptions = List (Input Abstract)

data TaskDescription where
  TaskDescription :: JsonTask -> InputDescriptions -> TaskDescription

instance ToJSON JsonTask

describe :: Members '[Alloc h, Read h] r => Task h t -> Sem r TaskDescription

data JsonInput where
    JsonInput :: Input Concrete -> JsonInput

instance FromJSON JsonInput
\end{lstlisting}

\paragraph{Visualize module}

In Listing \ref{sec:artefact:lst:visualize} we show the signatures of the
visualize module. We use this module to run the web server in production
(\texttt{visualizeTask}) or development (\texttt{visualizeTaskDevel}) mode. We
differentiate between these modes because we implemented live code reloading for
development, which requires a bit of additional setup. Both
\texttt{visualizeTask} and \texttt{visualizeTaskDevel} use the \texttt{initApp}
function. \texttt{InitApp} on its turn invokes the application-function of the
Application Module.

\begin{minipage}{\linewidth}
\begin{lstlisting}[caption={Visualize module (Haskell)},label={sec:artefact:lst:visualize},language=Haskell]
module Visualize (visualizeTask, visualizeTaskDevel) where

initApp :: ToJSON t => Task RealWorld t -> IO Application

visualizeTaskDevel :: ToJSON t => Task RealWorld t -> IO ()

visualizeTask :: ToJSON t => Task RealWorld t -> IO ()\end{lstlisting}
\end{minipage}
\subsection{Frontend}\label{sec:artefact:frontend}

The frontend renders the UI and provides a way for the user to interact with it. The code is written in PureScript using the Halogen framework.
The frontend consists of three main modules and some auxiliary modules. We
explain the main modules:

\begin{enumerate}
    \item The Client module is the communication layer with the backend. It
          defines functions which send requests to the backend and handles the
          responses.
    \item The Task module handles JSON encoding and decoding of our domain's
          datatypes (tasks and user input).
    \item The TaskLoader module is the starting point of Halogen and is
          responsible for rendering the UI.
\end{enumerate}

\paragraph{Client module}

The client module is responsible for the communication between frontend and
backend. The backend sends a response in JSON that consists of two parts: a
\texttt{Task} and a description of possible inputs. We decode this JSON object
into a \texttt{TaskResponse}. See Listing \ref{sec:artefact:lst:client}.

\begin{lstlisting}[caption={Client module (PureScript)},label={sec:artefact:lst:client},language=Haskell]
module App.Client (ApiError, TaskResponse(..), getInitialTask, interact, reset) where

data TaskResponse
    = TaskResponse Task (Array InputDescription)

instance decodeJsonTaskResponse :: DecodeJson TaskResponse

getInitialTask :: Aff (Either ApiError TaskResponse)

interact :: Input -> Aff (Either ApiError TaskResponse)

reset :: Aff (Either ApiError TaskResponse)\end{lstlisting}

\paragraph{Task module}
In the Client module we defined a \texttt{TaskResponse}. This
\texttt{TaskResponse} consists of two parts: a \texttt{Task} and an array of
\texttt{InputDescription}. In the Task module we define the decoding process of
\texttt{Task} and \texttt{InputDescription}. See Listing~\ref{sec:artefact:lst:task}.

\begin{minipage}{\linewidth}
\begin{lstlisting}[caption={Task module (PureScript)},label={sec:artefact:lst:task},language=Haskell]
module App.Task where

data Task
    = Edit Name Editor
    | Select Name Task Labels
    | Pair Task Task
    | Choose Task Task
    | Step Task
    | Trans Task
    | Done
    | Fail

instance showTask :: Show Task

instance decodeJsonTask :: DecodeJson Task

data Input
    = Insert Int Value
    | Decide Int String

instance showInput :: Show Input

instance encodeInput :: EncodeJson Input

data InputDescription
    = InsertDescription Int String
    | OptionDescription Int String

instance showInputDescription :: Show InputDescription

instance decodeJsonInputDescription :: DecodeJson InputDescription
    \end{lstlisting}
\end{minipage}

\paragraph{TaskLoader module}

The TaskLoader module renders the user interface (the \texttt{render} function
in Listing \ref{sec:artefact:lst:taskloader}). The module also contains logic to
handle events (\texttt{handleAction}), for example when a user modifies a value.
Finally, the \texttt{taskLoader} function (see Listing
\ref{sec:artefact:lst:taskloader}) initializes the component.

\begin{lstlisting}[caption={TaskLoader module (PureScript)},label={sec:artefact:lst:taskloader},language=Haskell]
module Component.TaskLoader (taskLoader) where

taskLoader :: forall query input output m. MonadAff m => H.Component query input output m

handleAction :: forall output m. MonadAff m => Action -> H.HalogenM State Action Slots output m Unit

render :: forall m. MonadAff m => State -> HH.ComponentHTML Action Slots m
\end{lstlisting}


\section{Examples}
\label{sec:examples}

We present a few examples to demonstrate how our framework handles TopHat programs.
The candy vending machine combines the Select and View editor, the
Step Task, and the Pair Task to construct a candy machine (Section~\ref{sec:example-candymachine}).
The calorie calculator demonstrates a real-world
application of our framework (Section~\ref{sec:example-calorie-calculator}). The chat sessions demonstrates the use of shared data stores (Section~\ref{sec:example-chat}), and finally Section~\ref{sec:tax} describes UI generation for the tax example from Section~\ref{sec:top:formal}

\subsection{Candy vending machine}
\label{sec:example-candymachine}

The candy machine allows a user to choose a chocolate bar and, after the bill is
paid, the candy machine returns the bar. The candy machine combines the Edit,
Pair and Step task. We have defined different Edit tasks with View and Select
editors. The implementation of the initial task is given in Listing
\ref{sec:artefact:lst:candymachine}. The Pair combinator is denoted with the
operator \lstinline{><}.

\begin{figure}[t]
    \subfloat[Step 1: Select a chocolate bar]{
        \includegraphics[width=\linewidth]{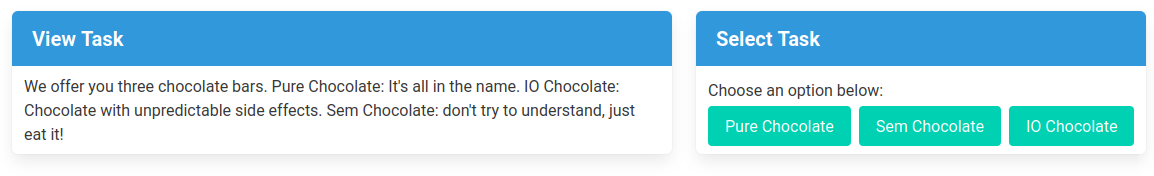}
        \label{fig:candymachine-step1}
    }

    \subfloat[Step 2: Price of the selected candy is shown to the user]{
        \includegraphics[width=\linewidth]{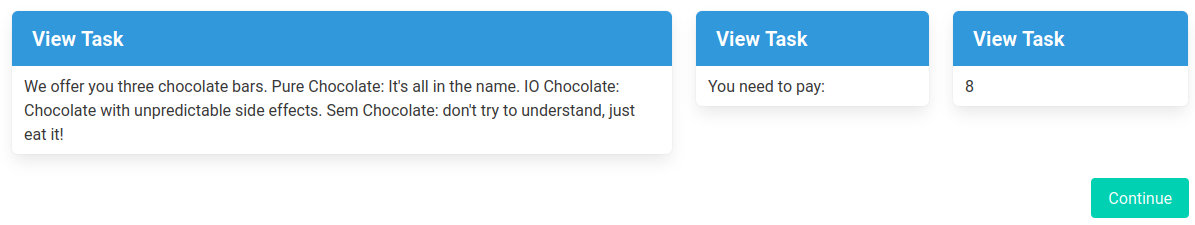}
        \label{fig:candymachine-step2}
    }

    \subfloat[Step 3: Insert a coin]{
        \includegraphics[width=\linewidth]{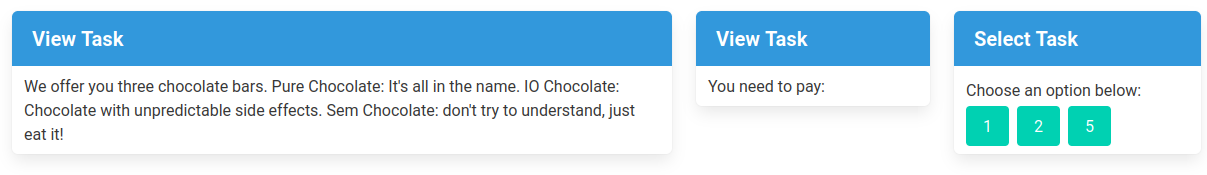}
        \label{fig:candymachine-step3}
    }

    \subfloat[Step 4: You have paid the bill]{
        \includegraphics[width=\linewidth]{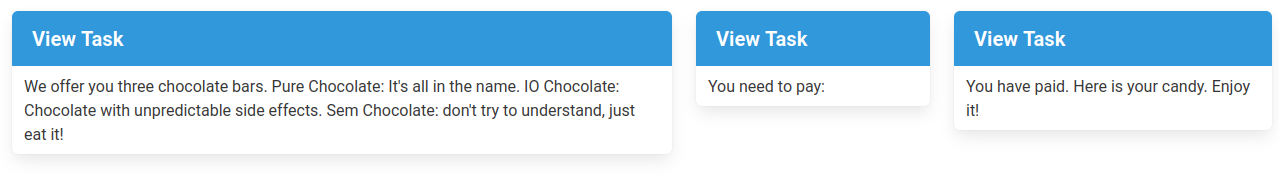}
        \label{fig:candymachine-step4}
    }
    \caption{Different stages of the candy vending machine}
\end{figure}

\begin{enumerate}
    \item After the candy machine is started, the machine displays some
          introductory text and a selection of chocolate bars (see Figure
          \ref{fig:candymachine-step1}). This is done using a Pair Task that consists
          of two Edit tasks: an Edit task with a View editor and an Edit Task with a
          Select editor.

    \item Select a chocolate bar. After choosing a bar, the candy machine
          displays the price of the bar (see Figure \ref{fig:candymachine-step2}).
          This is done using another Pair Task that consists of an Edit task with a
          View editor (\textit{``you need to pay:''}) and a Step Task. The Step task consists of
          two tasks: first a view editor is shown (with the price) and after the step,
          a select editor is rendered (see Figure \ref{fig:candymachine-step3}).

    \item Press the continue button.

    \item Insert coins until you have paid the bill (see Figure \ref{fig:candymachine-step3}). The application alternates a view and a select editor.

    \item The application shows a view editor to indicate to the user that the
          bill is paid (see Figure \ref{fig:candymachine-step4}).
\end{enumerate}


    \begin{lstlisting}[caption={Initial Task of the candy vending machine (Haskell)},label={sec:artefact:lst:candymachine},language=Haskell]
data CandyMachineMood = Fair | Evil

startCandyMachine :: (Task h (Text, (Text, Text)))
startCandyMachine = view "We offer you three chocolate
    bars. Pure Chocolate: It's all in the name. IO
    Chocolate: Chocolate with unpredictable side effects.
    Sem Chocolate: don't try to understand, just eat
    it!" >< select candyOptions

candyOptions :: HashMap Label (Task h (Text, Text))
candyOptions =
  [ entry "Pure Chocolate" 8,
    entry "IO Chocolate" 7,
    entry "Sem Chocolate" 9
  ]
  where
    entry :: Text -> Int -> (Label, Task h (Text, Text))
    entry name price =
      (name, view "You need to pay:" >< (view price >>? payCandy))

payCandy :: Int -> Task h Text
payCandy bill =
    select (payCoin bill) >>? \billLeft ->
        case compare billLeft 0 of
        EQ -> dispenseCandy Fair
        LT -> dispenseCandy Evil
        GT -> payCandy billLeft

payCoin :: Int -> HashMap Label (Task h Int)
payCoin bill =
  [ coinSize 5,
    coinSize 2,
    coinSize 1
  ]
  where
    coinSize :: Int -> (Label, Task h Int)
    coinSize size = (display size, view (bill - size))

dispenseCandy :: CandyMachineMood -> Task h Text
dispenseCandy Fair =
    view "You have paid. Here is your candy. Enjoy it!"
dispenseCandy Evil =
    view "You have paid too much! Sorry, no change, but here is your candy."
\end{lstlisting}


\subsection{Calorie calculator}
\label{sec:example-calorie-calculator}

To demonstrate a more real-world application that incorporates most task types,
we created a calorie calculator. This application calculates how many calories a
person should eat per day in order to maintain their weight. The calculation
depends on several factors, such as age, weight, and activity level. The
application can be broken down in several steps to prompt the user for input,
and finally calculating the result. The implementation of the task is
given in Listing \ref{sec:artefact:lst:calories}.

\begin{figure}
  \begin{center}
    \includegraphics[width=.7\linewidth]{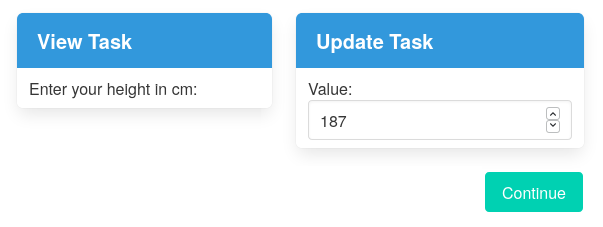}
  \end{center}
    \caption{Prompting the user to enter their height}
    \label{fig:calorie-calulator}
\end{figure}

\begin{enumerate}
    \item When started, the application presents the user with some information
          about the calculation using a View editor.
    \item After pressing continue, the user is prompted to enter the required
          data in different steps: height, weight, and age using Enter editors,
          and gender and activity level using Select editors. Each prompt is
          wrapped in a Pair task with a View editor on the left side to act as
          the label. Such a prompt is shown in
          Figure~\ref{fig:calorie-calulator}.
    \item In the last step the result is displayed using a View editor.
\end{enumerate}

\begin{figure}
\begin{lstlisting}[caption={Task of the calorie calculator (Haskell)},label={sec:artefact:lst:calories},language=Haskell]
data Gender = Male | Female

data ActivityLevel = Sedentary | Low | Active | VeryActive

type Height = Int

type Weight = Int

type Age = Int

calculateCaloriesTask :: Task h Text
calculateCaloriesTask =
    introduction >>? \_ -> do
        (_, height) <- promptHeight
        (_, weight) <- promptWeight
        (_, age) <- promptAge
        (_, gender) <- promptGender
        (_, activityLevel) <- promptActivityLevel
        let calories = calculateCalories gender activityLevel height weight age
        view
        ( "Your resting metabolic rate is: "
            <> display calories
            <> " calories per day."
        )

introduction :: Task h Text
introduction = view <| unlines
    [ "This tool estimates your resting metabolic rate,",
      "i.e. the number of  calories you have to consume",
      "per day to maintain your weight.",
      "Press \"Continue\" to start"
    ]

promptGender :: Task h (Text, Gender)
promptGender =
    view "Select your gender:"
        >< select
            [ "Male" ~> Done Male,
              "Female" ~> Done Female
            ]

promptHeight :: Task h (Text, Height)
promptHeight = view "Enter your height in cm:" >< enter

promptWeight :: Task h (Text, Weight)
promptWeight = view "Enter your weight in kg:" >< enter

promptAge :: Task h (Text, Age)
promptAge = view "Enter your age:" >< enter

promptActivityLevel :: Task h (Text, ActivityLevel)
promptActivityLevel =
    view "What is your activity level?"
        >< select
            [ "Sedentary" ~> Done Sedentary,
              "Low active" ~> Done Low,
              "Active" ~> Done Active,
              "Very Active" ~> Done VeryActive
            ]

-- We omit the actual calculation here since it is a bit lengthy.
calculateCalories :: Gender -> ActivityLevel -> Height -> Weight -> Age -> Int
calculateCalories gender al h w age = ...
\end{lstlisting}
\end{figure}

\subsection{Chat session}
\label{sec:example-chat}

\begin{figure}
    \centering
    \includegraphics[width=\linewidth]{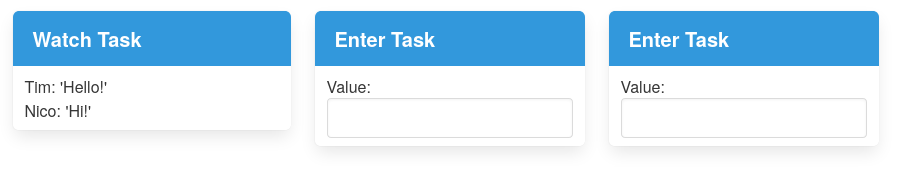}
    \captionsetup{type=figure}
    \captionof{figure}{A chat session using shared data stores.}
    \label{fig:tophat-chat-session}
\end{figure}

This example uses shared data stores to model a chat session between two
users, as displayed in Figure~\ref{fig:tophat-chat-session}. Each user can write
messages to the chat history on the left hand side using their respective inputs
on the right hand side.

The implementation for this example is given in Listing~\ref{lst:chat-session}.
The function \lstinline{share} creates a data store that can be accessed by
multiple tasks, in this case the two \lstinline{chat} tasks. The \lstinline{<<=}
operator is used to transform the contents of the shared data store.

\begin{minipage}{\linewidth}
\begin{lstlisting}[
    caption={A chat Session using shared data stores (Haskell)},
    label={lst:chat-session},
    language=Haskell]
chatSession :: Reflect h => Task h (Text, ((), ()))
chatSession = do
    history <- share ""
    watch history ><
        (chat "Tim" history >< chat "Nico" history)
    where
    chat :: Text -> Store h Text -> Task h ()
    chat name history = repeat <|
        enter >>* ["Send" ~> append history name]

    append :: Store h Text -> Text -> Text -> Task h ()
    append history name msg = do
        history <<= \h ->
        (if h == "" then h else h ++ "\n")
            ++ name ++ ": '"
            ++ msg ++ "'"
\end{lstlisting}
\end{minipage}

\subsection{Tax example}
\label{sec:tax}
For our final example, we revisit the tax program from Section~\ref{sec:top:formal}.

\begin{lstlisting}[caption={Tax example in Haskell},label={sec:tax:code},language=Haskell]
tax :: Task h ((((Amount, Bool), Bool), Date), Date)
tax =
  let today :: Date
      today = 100

      provideDocuments :: Task h (Amount, Date)
      provideDocuments = enter >< enter

      companyConfirm :: Task h Bool
      companyConfirm = enter

      officerApprove :: Date -> Date -> Bool -> Task h Bool
      officerApprove invoiceDate date confirmed =
        view (date - invoiceDate < 365 && confirmed)
   in (provideDocuments >< companyConfirm)
        >>? \((invoiceAmount, invoiceDate), confirmed) ->
          officerApprove invoiceDate today confirmed
            >>? \approved ->
              let subsidyAmount =
                   if approved
                      then min 600 (invoiceAmount `div` 10)
                      else 0
                      in view
             <| unlines
               [ "Subsidy amount: " ++ display subsidyAmount,
                 "Approved: " ++ display approved,
                 "Confirmed: " ++ display confirmed,
                 "Invoice date: " ++ display invoiceDate,
                 "Today: " ++ display today
               ]
\end{lstlisting}

Listing~\ref{sec:tax:code} gives the Haskell code that implements the task.
Compared to the original definition as given in Listing~\ref{lst:tax}, the task is nearly identical.
The only change made is to the final line, where we have opted for a different presentation of the final result, for simplicity's sake.

\begin{figure}[t]
    \subfloat[Step 1: The citizen enters the request info on the left, the installation company confirms on the right]{
        \includegraphics[width=\linewidth]{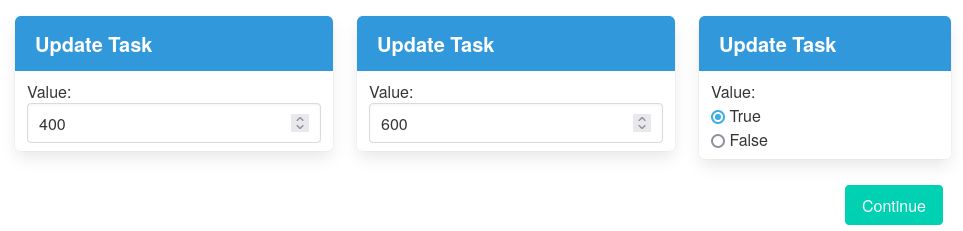}
        \label{}
    }

    \subfloat[Step 2: The tax office confirms or denies the request]{
    \hspace{.5cm}
    \includegraphics[width=.30\linewidth]{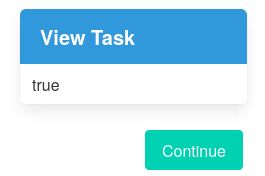}
        \label{}
    }\hspace{1cm}
    \subfloat[Step 3: The final outcome of the request is displayed]{
        \includegraphics[width=.30\linewidth]{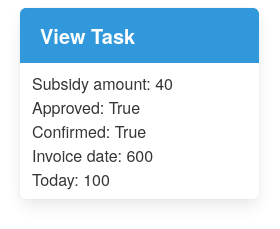}
    }
    \caption{Different stages of the tax subsidy application}
    \label{fig:tax}
\end{figure}

Figure~\ref{fig:tax} lists the different stages of the UI for the tax subsidy task.
First, the user requesting the subsidy can enter in information (first two tasks), while the company can confirm or deny.
Then, the tax officer can verify if the conditions are met, and approve the request.
Finally, the outcome is shown.

Since we did not have to modify the task at all, besides a minor presentation detail, this task can still be proven correct using symbolic execution.
This example clearly illustrates the advantage of TopHat with a UI over the current state-of-the-art in the form of iTasks.


\section{Related work}
\label{sec:related-work}

Section~\ref{sec:top} presentend related work on TOP and iTasks.
In this section, we will briefly discuss Functional reactive programming as an alternative to TOP, as well as alternatives for the UI framework and web server we have used during the development of the UI for TopHat.


\subsection{Functional Reactive Programming}

Functional Reactive programming (FRP) is another approach to UI development
using functional programming. FRP is a programming paradigm centered around
interactive event-based applications. It has implementations in multiple
programming languages, such as Haskell and
JavaScript~\cite{Bainomugisha2013reactive}.

FRP consists of two main concepts: \textit{behaviors} and \textit{events}. A
behavior consists of a value and can be mapped to output, for example a label.
Behaviors can depend on other behaviors, so a change in a behavior can propagate
through a network of dependent behaviors. An event only occurs at a certain
point in time and contains a value. Input is mapped to events, for example the
pressing of a key or the position of the mouse cursor. Events can trigger
changes in behaviors.

It is worth noting that, while they share some similarities, FRP and TOP are
conceptually different. FRP is a paradigm for reactive programming, whereas TOP
is a way to model collaboration between users.

\subsection{User Interface frameworks}

We build upon the Halogen framework to create our prototype, but many other UI
frameworks exist in the domain of functional programming.
We discuss three of these briefly below.

Elm~\cite{czaplicki2012elm} refers to both Elm, a functional programming
language that compiles to JavasScript~\cite{elmlang}, and
TEA~\cite{elmarchitecture}, a programming pattern that emerged from it.
Elm's ecosystem consists of a large number of available libraries that help in
creating web applications.

%


Miso~\cite{haskellmiso} is a Haskell front-end framework inspired by Elm and
Redux. It relies on GHCJS~\cite{ghcjs}, a Haskell-to-JavaScript compiler based
on GHC.

%
%


Reflex~\cite{reflexfrp} is an FRP framework written in Haskell with support for
a variety of platforms, including the web, desktop, and mobile. Reflex
applications are modular, which makes growing and refactoring an application
efficient and swift.

%
%


We have selected PureScript and Halogen because it is a powerful functional programming language
that fits our problem domain. Halogen provides an excellent developer
experience, has a component based architecture and builds upon PureScript's
power and expressiveness.

\subsection{Web servers}

We have opted for Servant as our web server.
Servant provides combinators to implement our features, which makes
coding less error prone and time-consuming. Servant is up-to-date,
well-maintained, well documented and it is easy to get a working prototype.
Below we discuss Yesod and Warp as possible alternatives for the server used in our implementation.

The Yesod Web Framework~\cite{snoyman} is a Haskell web framework that allows for rapid development of type-safe, RESTful and high performance web applications~\cite{yesod}.
The Yesod Web Framework adds the strengths of Haskell (like type safety) to the web.
Especially on the boundaries of Yesod and the world, for example a user enters input or persistent data is loaded, Yesod adds mechanisms to define the expected types~\cite{yesodBook}. We found that developing a prototype based on Yesod is more difficult than developing a prototype based Servant. We also found that the Yesod Web Framework is too extensive for our purposes~\cite{markmarc2021}.



The Warp web server is a light-weight web server that supports the Web Application Interface (WAI)~\cite{snoyman2011warp}.
It is meant to be easy to use and provide easy composition of web services.
Because of the design choices to achieve this, the code of a Warp prototype is low-level.
This means that implementing all features in this way will be error prone and time-consuming.
Therefore, we have chosen Servant. However, Servant also uses Warp as its web server~\cite{markmarc2021}.



\section{Conclusion}
\label{sec:conclusion}

We have demonstrated TopHat UI, a proof-of-concept framework that implements a GUI for Tophat programs.
None of the advanced Clean features used by iTasks were required to do so, as expected.
On top of that, we were able to leave the TopHat language untouched, preserving its formal properties.

Our framework implements all basic requirements for a TOP framework, by supporting tasks, shared data stores, combinators and generics.
The source code for our framework is available online, and can thus be leveraged by developers and researchers to advance the field of Task-Oriented Programming.

\subsection{Future work}

As mentioned in Section~\ref{sec:tophat-user-interface}, TOP features such as multi-user support and richer datatypes are considered future work.
We see no technical or formal reason prohibiting them from being included in future versions of the UI framework.
As with iTasks, the rendering of values, and editors of values, is generic in the type of the value.
Adding support for more complex datatypes would just mean making instances for them for viewing and editing them, similar to how this is done in iTasks.
As for multi-user support, this is a limitation in the current version of TopHat.
Its developers are already working on adding multi-user support.
Once this feature is released, we see no fundamental limitations in supporting this in the UI.
The server framework used in the current implementation, Servant, already has extensive support for user authentication, which could be leveraged~\footnote{\url{https://docs.servant.dev/en/stable/tutorial/Authentication.html}}.



\bibliographystyle{ACM-Reference-Format}
\bibliography{main}

\end{document}